\documentclass[lettersize,journal]{IEEEtran}
\usepackage{amsmath,amsfonts}
\usepackage{algorithmic}
\usepackage{algorithm}
\usepackage{array}
\usepackage[caption=false,font=normalsize,labelfont=sf,textfont=sf]{subfig}
\usepackage{textcomp}
\usepackage{stfloats}
\usepackage{url}
\usepackage{verbatim}
\usepackage{graphicx}
\usepackage{cite}

\begin{document}

\title{Public Perceptions of Autonomous Vehicles: A Survey of Pedestrians and Cyclists in Pittsburgh}

\author{Rudra Yashodhan Bedekar\\
George Mason University}

\maketitle

\begin{abstract}
This study investigates how autonomous vehicle (AV) technology is perceived by pedestrians and bicyclists in Pittsburgh. Using survey data from over 1200 respondents, the research explores the interplay between demographics, AV interactions, infrastructural readiness, safety perceptions, and trust. Findings highlight demographic divides, infrastructure gaps, and the crucial role of communication and education in AV adoption.
\end{abstract}


\section{Introduction}
Autonomous vehicle (AV) integration into urban settings has sparked serious concerns about how these vehicles may affect vulnerable road users, especially pedestrians and cyclists. It is critical to comprehend the comfort, safety, and views of these road users as autonomous vehicles (AVs) are tested and used more frequently in places like Pittsburgh.\\
Sharing the road with autonomous vehicles poses special risks for pedestrians and cyclists because of their exposure and lack of physical protection. Among these issues are worries regarding AVs' capacity to recognize and react to their motions, especially in situations with a lot of traffic or unpredictability. Furthermore, concerns and discomfort may be exacerbated by the inadequacy of the current urban infrastructure to facilitate the safe coexistence of AVs and non-motorized users.\\
Finding the elements that affect how pedestrians and cyclists view AVs, the degree to which demographic characteristics and prior experiences influence these opinions, and the typical difficulties or anxieties encountered while interacting with AVs constitute the main research challenge. In order to guarantee the safe and fair implementation of AV technology in urban settings, this issue must be resolved. This study intends to offer practical insights into how AVs might be incorporated into current transportation networks without endangering the comfort and safety of vulnerable road users by concentrating on Pittsburgh, a city at the forefront of AV testing.

\section{Importance}
Given that bicycles and pedestrians are vulnerable road users with no physical protection,
it is imperative to comprehend their attitudes and worries about autonomous vehicles
(AVs). The way AVs maneuver in public areas like bike lanes, crosswalks, and junctions
has a direct effect on their comfort and safety. In order to reduce accidents, increase public
confidence in AV technology, and guarantee its inclusive and equitable incorporation
into urban settings, these problems must be resolved. The deployment of AVs runs
the potential of escalating safety concerns and deterring active transportation, such as
walking and cycling, which are essential for sustainable urban mobility, if their viewpoints
are not taken into account.\\
Understanding the attitudes and concerns of pedestrians and cyclists regarding au-
tonomous vehicles (AVs) is crucial since they are vulnerable road users who lack physical
protection. The comfort and safety of AVs are directly impacted by how they navigate
public spaces including bike lanes, crosswalks, and intersections. These issues need to be
fixed in order to lower accident rates, boost public trust in AV technology, and ensure its
fair and inclusive integration into urban environments. If their opinions are ignored, the
deployment of AVs could increase safety concerns and discourage active transportation,
including walking and bicycling, which are crucial for sustainable urban mobility.

\section{Research Questions}
\begin{enumerate}
\item What aspects affect the comfort or discomfort level of Pittsburgh's bicycles and pedestrians when it comes to sharing the road with autonomous vehicles? \\      1Perceptions of autonomous vehicles are significantly influenced by demographic factors like age and previous riding or walking experience. Policymakers and AV developers can better grasp the differing comfort levels and concerns of various demographic groups by looking at these characteristics, which will allow for more inclusive and focused outreach campaigns.

\item What impact do demographic variables (age and prior experience riding or walking) have on people's perceptions of autonomous cars? \\ 
Age and previous walking or bicycle experience are two demographic factors that might greatly influence how people view autonomous vehicles. Policymakers and city planners can gain a better understanding of how various groups respond to AV technology by examining these characteristics. This promotes trust in AV systems and ensures that infrastructure construction and public engagement are inclusive, accommodating a range of viewpoints.

\item  To what extent may contextual and demographic characteristics be used to predict the perceived safety of autonomous vehicles? \\
The study reveals significant disparities in Pittsburgh's age and gender groups' familiarity with technology and frequency of interactions with autonomous vehicles (AVs). Comparing younger and older age groups, the former indicate higher levels of technology familiarity and more frequent contacts with AVs. In addition, across all age groups, men are more likely than women to be more acquainted with technology and come into contact with AVs. According to these results, demographic variables—specifically, age and gender—have a big impact on how people use and feel about AV technology.

\item What is the relationship between reported safety incidents and familiarity with news regarding autonomous vehicles and the readiness to provide trip data? \\
According to the investigation, there is a significant correlation between the readiness to share trip data, familiarity with news regarding autonomous vehicles (AVs), and reported safety problems. Participants are often more eager to contribute their trip data if they are more familiar with AV-related news, suggesting that awareness is a key factor in building participation and confidence. On the other hand, people who have less knowledge or experience with AV issues are more likely to be hesitant or unsure about disclosing their information. This implies that raising public knowledge and educating people about AV technology and its advantages may improve trust and promote wider involvement in data-sharing programs that are crucial to the development of AV systems.

\item Does having a car AutOwner and being tech-savvy FamiliarityTech affect people's perceptions of how safe autonomous vehicles SafeAv are? Does the age of the individual affect this opinion as well? \\
The purpose of this study is to investigate whether people's opinions about autonomous vehicle (AV) safety are influenced by their automobile ownership status and level of technological familiarity. It also looks into whether these opinions differ depending on the age group. By examining these variables, the study aims to comprehend how attitudes toward AV safety, technological familiarity, and demographics relate to one another.

\item How do perceptions of safety among pedestrians and cyclists vary by the type of autonomous vehicle interaction they experience? \\ Analyzing and visualizing the effects of varying degrees of AV interaction on non-motorized road users' sense of safety is the goal. Three categories of contacts are distinguished by the study: close meetings with no incidents, genuine AV-related incidents, and simple observations of AVs. In order to provide a clear visual depiction of how direct or indirect experiences with AVs affect people's sense of safety, the average safety perception score for each category will be shown in a bar chart. In addition to highlighting the differences in safety perceptions according on the type of interaction, this method helps identify possible places for AV technology and urban design advancements to promote public safety and trust.

\end{enumerate}

\section{Literature Review}
This report analyzes the effects of autonomous vehicles (AVs) on traffic safety and transportation systems, emphasizing their potential to enhance traffic flow, decrease accidents, and transform urban mobility. The results indicate that the predictability of AV behavior and efficient communication are essential for ensuring the comfort and safety of non-drivers, especially bicyclists and pedestrians. Elements such as the behavior of AVs, their predictability, and communication significantly influence the comfort levels of road users, while demographic factors like age, gender, and previous experiences shape perceptions of AVs. Generally, younger individuals exhibit greater comfort with AVs, whereas older adults and women often voice increased concerns regarding personal safety. Cyclists and pedestrians commonly express apprehensions about AVs' capacity to react to sudden movements, the lack of human-like interaction signals, perceived aggressiveness, and the silent operation of AVs, which reduces auditory awareness. These findings highlight the necessity for enhanced design and communication strategies to facilitate effective interactions between AVs and non-drivers, thereby fostering public trust. \\
This report examines the application of virtual reality simulations to investigate the communication dynamics between pedestrians and autonomous vehicles (AVs). It underscores the necessity for clear and comprehensible signals to enhance safety and build trust. The results indicate that the behavior, predictability, and communication of AVs play a crucial role in influencing the comfort levels of non-drivers, consistent with studies conducted in Pittsburgh focusing on cyclists and pedestrians. Key factors such as predictable vehicle actions, distinct visual or auditory signals, and suitable AV speeds contribute to safety and alleviate anxiety, whereas abrupt movements and unclear signaling tend to heighten discomfort. Infrastructure components, including dedicated bike lanes, prominent signage, and sufficient buffer zones, are essential in fostering a sense of security. Furthermore, demographic variables such as age, gender, and previous experiences affect perceptions of AVs, with younger individuals generally exhibiting greater comfort with the technology than older adults, while women often express heightened concerns regarding personal space and safety. Cyclists and pedestrians consistently voice safety apprehensions, particularly regarding AVs' capacity to recognize unexpected movements, their silent operation diminishing auditory awareness, and uncertainties surrounding right-of-way. These insights highlight the critical need for well-structured AV communication systems and infrastructure to enhance the comfort and trust of non-drivers in urban environments. \\
This report examines the application of virtual reality simulations to investigate the communication dynamics between pedestrians and autonomous vehicles (AVs). It underscores the necessity for clear and comprehensible signals to enhance safety and build trust. The results indicate that the behavior, predictability, and communication of AVs play a crucial role in influencing the comfort levels of non-drivers, consistent with studies conducted in Pittsburgh focusing on cyclists and pedestrians. Key factors such as predictable vehicle actions, distinct visual or auditory signals, and suitable AV speeds contribute to safety and alleviate anxiety, whereas abrupt movements and unclear signaling tend to heighten discomfort. Infrastructure components, including dedicated bike lanes, prominent signage, and sufficient buffer zones, are essential in fostering a sense of security. Furthermore, demographic variables such as age, gender, and previous experiences affect perceptions of AVs, with younger individuals generally exhibiting greater comfort with the technology than older adults, while women often express heightened concerns regarding personal space and safety. Cyclists and pedestrians consistently voice safety apprehensions, particularly regarding AVs' capacity to recognize unexpected movements, their silent operation diminishing auditory awareness, and uncertainties surrounding right-of-way. These insights highlight the critical need for well-structured AV communication systems and infrastructure to enhance the comfort and trust of non-drivers in urban environments.

\section{Dataset Description}
The Pittsburgh Department of Transportation initially gathered the "Autonomous Vehicle Survey of Bicyclists and Pedestrians in Pittsburgh" dataset, which was then used in this study's secondary analysis. Data.gov, a public database of government information, provided access to the dataset (Pittsburgh Department of Transportation, 2021). Other than giving due credit and acknowledgment to the original source, no extra permissions were needed to utilize this publicly available open data in this study.

In order to learn more about how the public views autonomous cars, the Pittsburgh Department of Transportation conducted a survey from March to June 2021 as part of the city's Smart City effort. With an emphasis on vulnerable road users including bicycles and pedestrians, the original data collection's main goal was to assist city planning and policy decisions about the integration of autonomous vehicles in urban areas.

The dataset includes survey answers from 1200+ people who lived in the Pittsburgh metropolitan region and identified as either cyclists, pedestrians, or both. Six age categories were represented in the sample demographically: 18–24, 25–34, 35–44, 45–54, 55–64, and 65+. Of the respondents, roughly 58\% belonged to the 25–34 and 35–44 age groups. The distribution of genders was roughly 46\% male, 52\% female, and 2\% non-binary or wanting to remain anonymous. 47 distinct Pittsburgh zip codes were included in the geographic coverage, with downtown and university regions having the largest presence.

The dataset provides a thorough examination of the opinions and experiences of the general population about autonomous vehicles (AVs) among those who do not drive motorized vehicles. It collects input from cyclists and pedestrians, highlighting the key factors influencing their perceptions of safety, comfort, and trust when using shared roads with autonomous vehicles. Taking into account factors including vehicle speed, predictability, and communication tactics, participants evaluated how safe and comfortable they felt interacting with AVs. Along with evaluating confidence levels based on exposure to and experience with AVs, the information also highlights preferences for different road infrastructures, such as shared lanes, bike lanes, and pedestrian crossings.

Several problems were found during the data quality assessment and processing phase: outliers were found in around 2.7\% of the survey replies, and missing values were found in about 4.2\% of the responses, mostly in open-ended parts. There were also other category factors that showed inconsistencies. These were addressed by standardizing category labels, removing responses with more than 30\% missing data, handling outliers with visualization approaches, and using mean imputation for missing numerical values. In terms of ethics, the study used de-identified, publicly accessible data that complied with APA standards and did not include any personal information. An institutional research board approved the study, guaranteeing that it complied with ethical guidelines for the processing of secondary data.

\section{Methods and Tools}
During the data cleaning process, the pandas package in Python was used to format the dataset, manage missing values, and correct errors. Both R and Python were used for exploratory data analysis (EDA) and advanced analytics. Trend analysis, descriptive statistics, and visualizations such as scatter plots and histograms were made possible with Python's pandas, matplotlib, and seaborn modules. Plots and statistical summaries of publication quality were added to these visualizations using R's ggplot2 and dplyr tools. R's caret packages and Python's scikit-learn were used for regression analysis and predictive modeling. AWS services like EC2 were used to run all scripts in a scalable cloud environment, guaranteeing effective processing and analysis. Combining AWS Glue, Athena, SQL, Python, and R gave researchers strong tools to completely meet the study's goals and extract insightful information.
Several tools and methods were used to effectively prepare, analyze, and interpret the dataset for this research. Amazon S3, a central storage platform, made that the data was expandable, safely stored, and quickly accessible. The data was arranged and prepared for analysis using AWS Glue, which facilitates the processing of big datasets. To obtain particular subsets of data straight from S3 for in-depth analysis and aggregation, SQL queries were executed using Amazon Athena. This eliminated the need for manual data movement or preprocessing, enabling effective querying and analysis.

\section{Results and Findings}
\subsection{Data Exploration Using Python and R} 
\begin{figure}[h!]
    \centering
    \includegraphics[width=8cm]{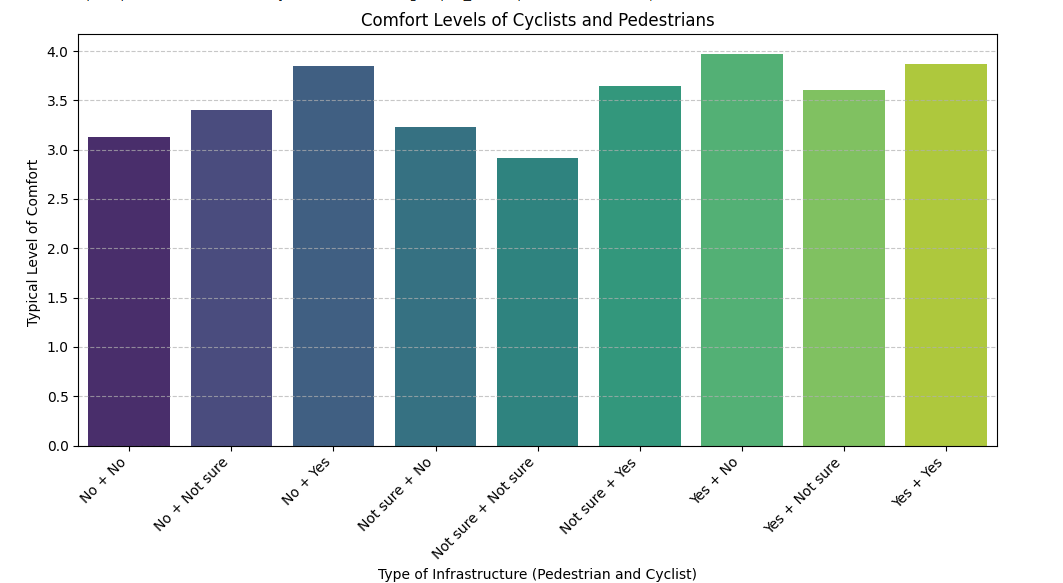}
    \caption{Comfort Levels for Shared Infrastructure Combinations}
    \label{fig:q1}
\end{figure}

Under different infrastructural combinations, the bar chart (Figure \ref{fig:q1}) illustrates how comfortable Pittsburgh's bicycles and pedestrians are sharing the road with autonomous vehicles (AVs). According to the research, people report feeling most comfortable when there is infrastructure for both bicyclists and pedestrians ("Yes + Yes"), highlighting the significance of designated routes for both populations. Relatively high comfort levels are also demonstrated by infrastructure combinations like "Yes + No" and "No + Yes," when at least one type of shared infrastructure is accessible. On the other hand, combinations like "Not sure + No" and "Not sure + Not sure" lead to lower levels of comfort, indicating that perceived safety and comfort are greatly reduced by uncertainty or the lack of specialized infrastructure.\\

The association between age, cycling experience, and people's opinions about autonomous vehicle (AV) safety across age groups is depicted in the bar chart (Figure \ref{fig:q2}). Across all age groups, those with shared cycling experience (shown by the orange bars) consistently report higher perceptions of AV safety than those without (shown by the blue bars). Overall, younger participants—especially those between the ages of 18 and 24—have the highest safety perceptions, and their comfort levels are further enhanced by prior cycling experience. Older individuals, such as those 65 and older, on the other hand, have somewhat lower safety perceptions, though their prior riding experience still has a positive impact. This investigation highlights how perceptions of AV safety are significantly influenced by both age and cycling experience. Confidence in AV safety is higher among younger people, who are perhaps more tech-savvy and accustomed to the dynamics of contemporary road-sharing. Increased exposure to shared infrastructure and AV interactions may be the reason why shared cycling experience seems to close comfort disparities across all age groups. \\

\begin{figure}[h!]
    \centering
    \includegraphics[width=8cm]{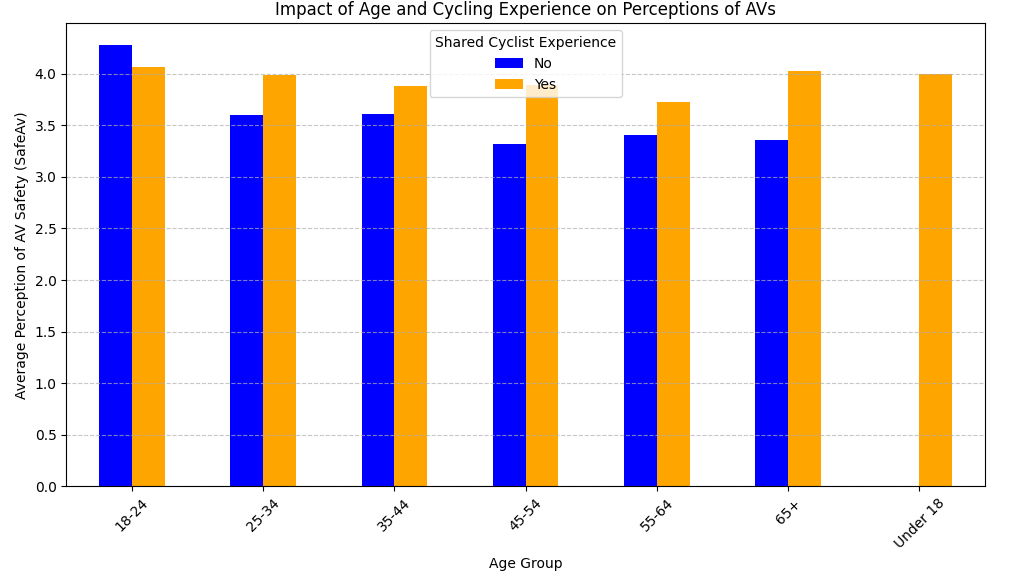}
    \caption{Impact of Age and Cycling Experience on AV Safety Perceptions}
    \label{fig:q2}
\end{figure}

The information sheds light on how pedestrians and bikers in Pittsburgh view and engage with autonomous vehicles (AVs). Demographics (e.g., age, zip code), knowledge of AV-related news and technology, opinions on how safe AVs are in comparison to human drivers, and readiness to share trip and performance data are some of the criteria it considers. The attitudes and experiences of the respondents are shown by important fields such as AVImpact, FamiliarityTech, and SafeAv (perception of AV safety). Furthermore, information on smartphone and bicycle ownership, collision perceptions, and the use of shared infrastructure provides a thorough understanding of the variables affecting AV comfort and confidence.

\begin{figure}[h!]
    \centering
    \includegraphics[width=\linewidth]{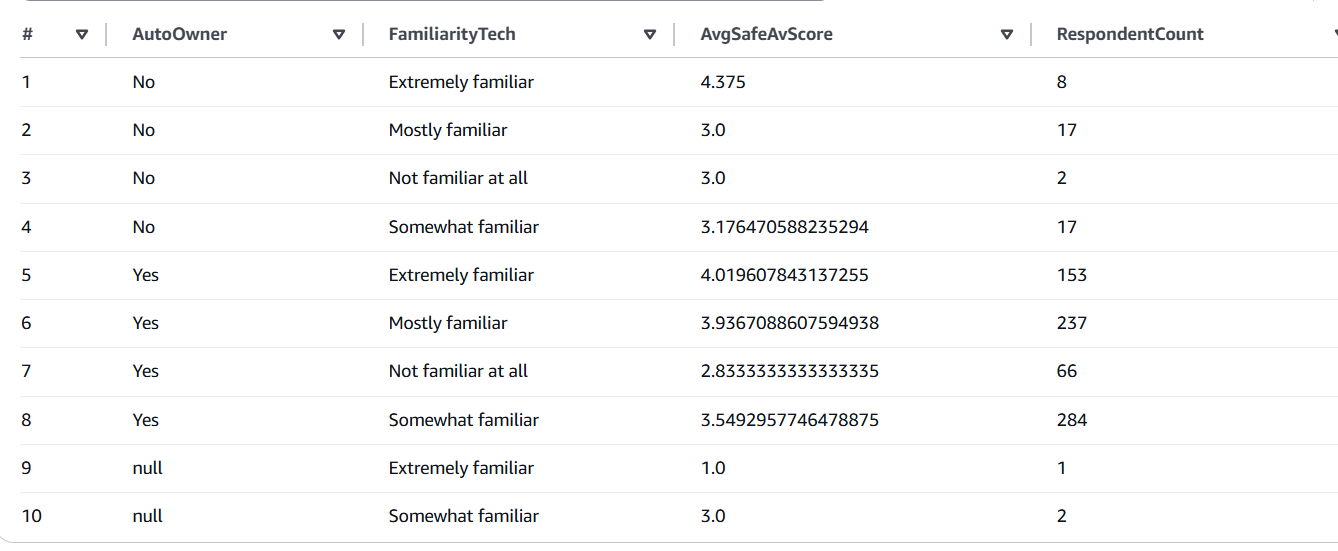}
    \caption{Overview of the dataset: AV perceptions, technological familiarity, and demographics}
    \label{fig:q2}
\end{figure}

On a scale of 1 to 5, the graph (Figure \ref{fig:q3}) contrasts observed and expected values for AV safety perceptions. Actual participant-reported safety perceptions are represented by observed values, whereas a regression model that takes contextual and demographic factors into account produces predicted values. Although there is some agreement between the model's predicted and actual values, significant differences imply that the regression does not account for all the variables affecting safety perceptions.\\

\begin{figure}[h!]
    \centering
    \includegraphics[width=8cm]{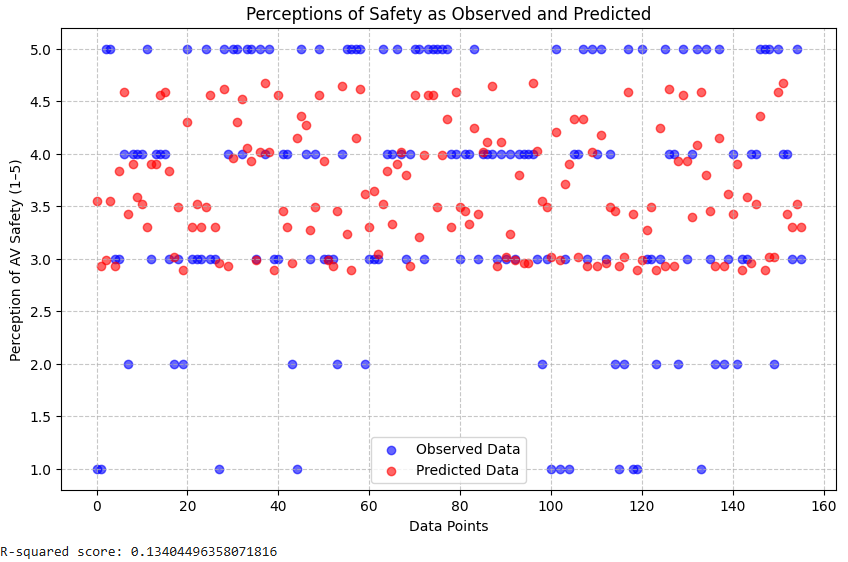}
    \caption{Observed vs. Predicted Safety Perceptions for AVs}
    \label{fig:q3}
\end{figure}

The average squared difference between the expected and actual values of AV safety perceptions  is indicated by the Mean Squared Error (MSE) of 1.227. Better model performance is indicated by a lower MSE, where predictions are more in line with the actual data. To properly evaluate the correctness of the model, additional analysis (such as R-squared) may be useful, as this value also depends on the scale of the target variable 

\begin{figure}[h!]
    \centering
    \includegraphics[width=8cm]{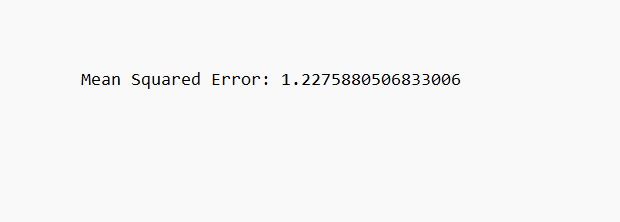}
    \caption{Mean Squared Error}
    \label{fig:q3}
\end{figure}

\begin{figure}[h!]
    \centering
    \includegraphics[width=8cm]{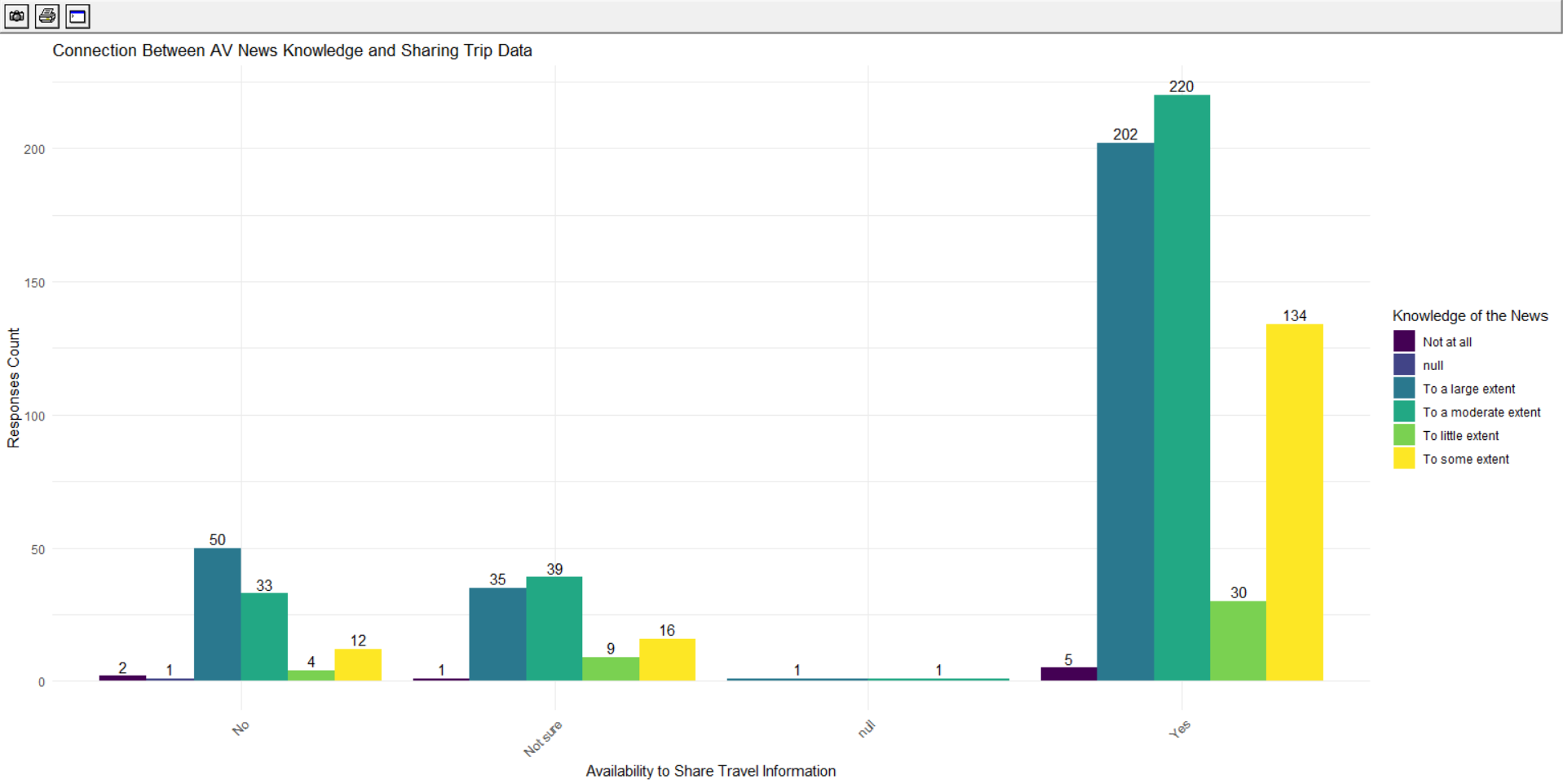}
    \caption{Relationship Between AV News Familiarity and Willingness to Share Trip Data}
    \label{fig:q4}
\end{figure}

The relationship between people's willingness to share trip data and their familiarity with news regarding autonomous vehicles (AVs) is depicted in the bar chart (Figure \ref{fig:q4}). Responses are categorized according to whether they choose to share travel information ("Yes," "No," or "Not sure") and their degree of acquaintance with AV news ("Not at all," "To a large extent," etc.). Those who answered "To a large extent" or "To a moderate extent," which indicate greater familiarity with AV news, are more likely to reply "Yes" to sharing trip data, according to the graphic. Conversely, people who are unfamiliar with the question ("Not at all" or "To little extent") are more likely to select "No" or "Not sure."\\

\subsection{Data interpretation using SQL in AWS-Athena}
The table looks at people who don't own cars and demonstrates how age and technological familiarity affect how safe people think autonomous cars are. AVs are often rated as safer by respondents with increased familiarity, such as "Extremely familiar," particularly younger groups (18–24, 25–34). On the other hand, especially among older respondents, lower familiarity levels are associated with lower safety assessments. This demonstrates how age and technology familiarity affect perceptions of AV safety.

\begin{verbatim}
SELECT AutoOwner, FamiliarityTech, Age, 
AVG(SafeAv)
AS AvgSafeAvScore, 
COUNT(*) AS RespondentCount
FROM projectaws777
WHERE SafeAv IS NOT NULL
AND Age IS NOT NULL
GROUP BY AutoOwner, FamiliarityTech, Age
ORDER BY AutoOwner, FamiliarityTech, Age;
\end{verbatim}

\begin{figure}[h!]
    \centering
    \includegraphics[width=8cm]{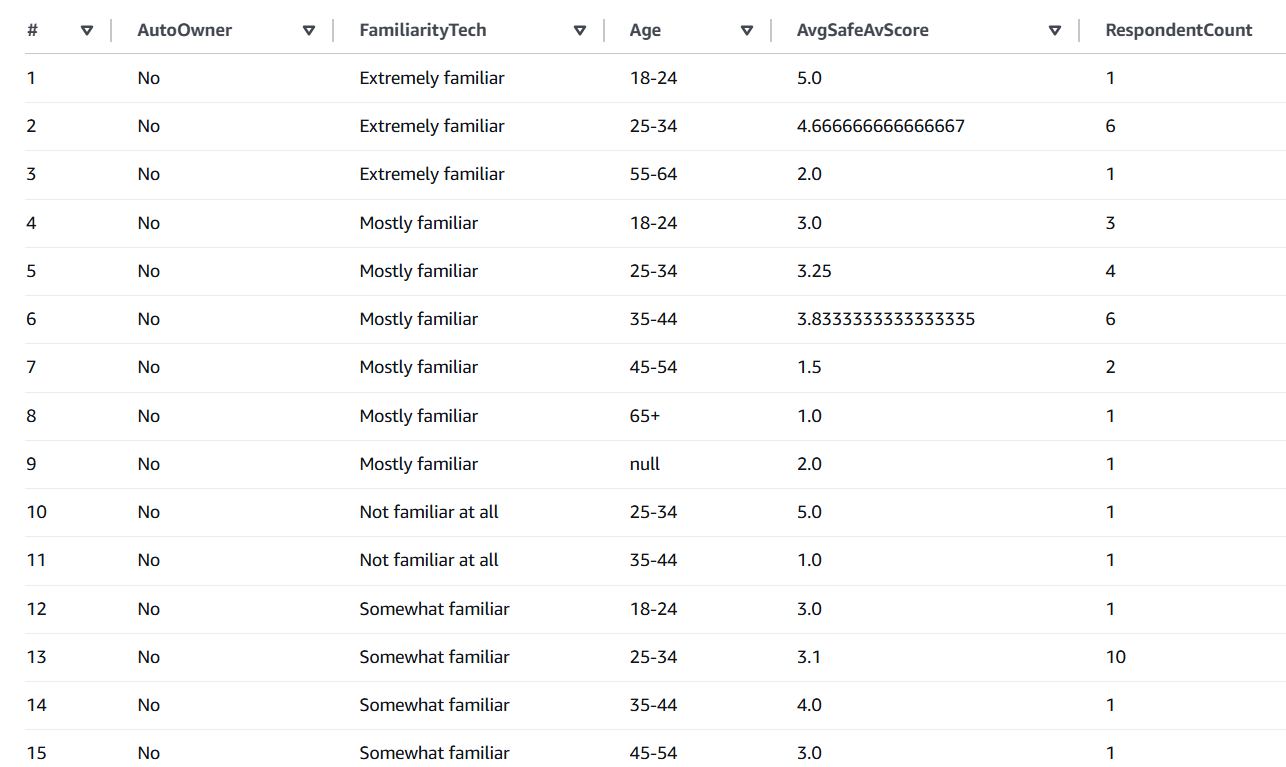}
    \caption{
Average ratings for AV safety by age group}
    \label{fig:q5}
\end{figure}

The average perceived safety of autonomous vehicles (AvgSafeAvScore) is influenced by age groups, as the table illustrates. Middle-aged and older groups score slightly lower than younger groups, with those aged 18 to 24 giving AVs the highest rating (4.09), while those aged 55 to 64 give them the lowest rating (3.49). The respondents with the lowest average score (2.86) are those who did not specify their age. This demonstrates a pattern where younger people believe AVs to be safer than older groups.

\begin{verbatim}
SELECT  Age, AVG(SafeAv) AS AvgSafeAvScore, 
    COUNT(*) AS RespondentCount
FROM projectaws777
WHERE SafeAv IS NOT NULL AND Age IS NOT NULL
GROUP BY  Age
ORDER BY Age;
\end{verbatim}

\begin{figure}[h!]
    \centering
    \includegraphics[width=8cm]{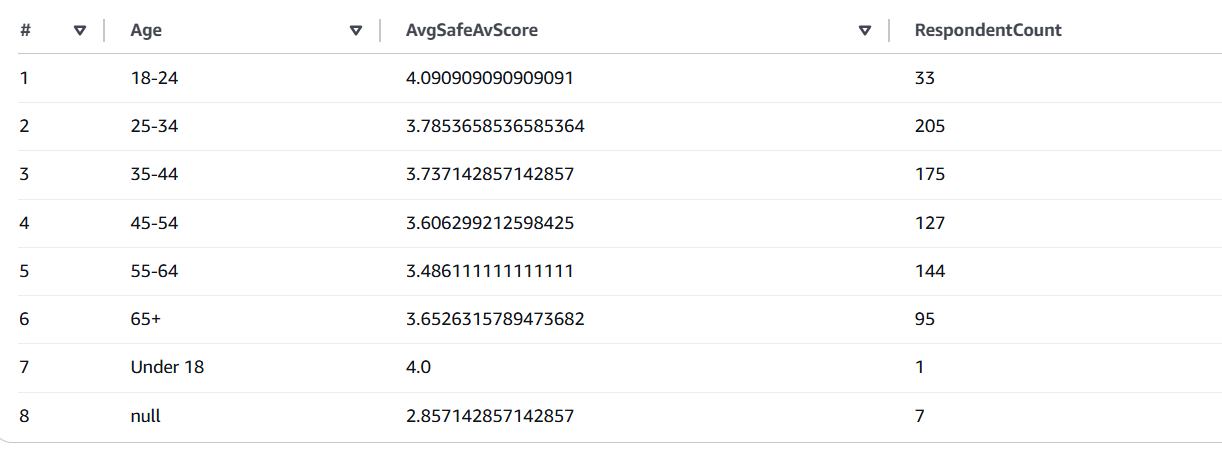}
    \caption{AV safety ratings by age group}
    \label{fig:q5}
\end{figure}

According to vehicle ownership AutoOwner and technological familiarity FamiliarityTech, the average perceived safety scores of autonomous vehicles AvgSafeAvScore are compared in the table. Even if they do not own a car, those who are "Extremely familiar" with technology tend to give AVs better ratings. On average, though, car owners score marginally higher on safety than non-owners across the majority of acquaintance levels. Each category's participation distribution is reflected in the respondent count RespondentCount.

\begin{verbatim}
SELECT  AutoOwner, FamiliarityTech, 
    AVG(SafeAv) AS AvgSafeAvScore, 
    COUNT(*) AS RespondentCount
FROM projectaws777
WHERE SafeAv IS NOT NULL
GROUP BY AutoOwner, FamiliarityTech
ORDER BY AutoOwner, FamiliarityTech;
\end{verbatim}

\begin{figure}[h!]
    \centering
    \includegraphics[width=8cm]{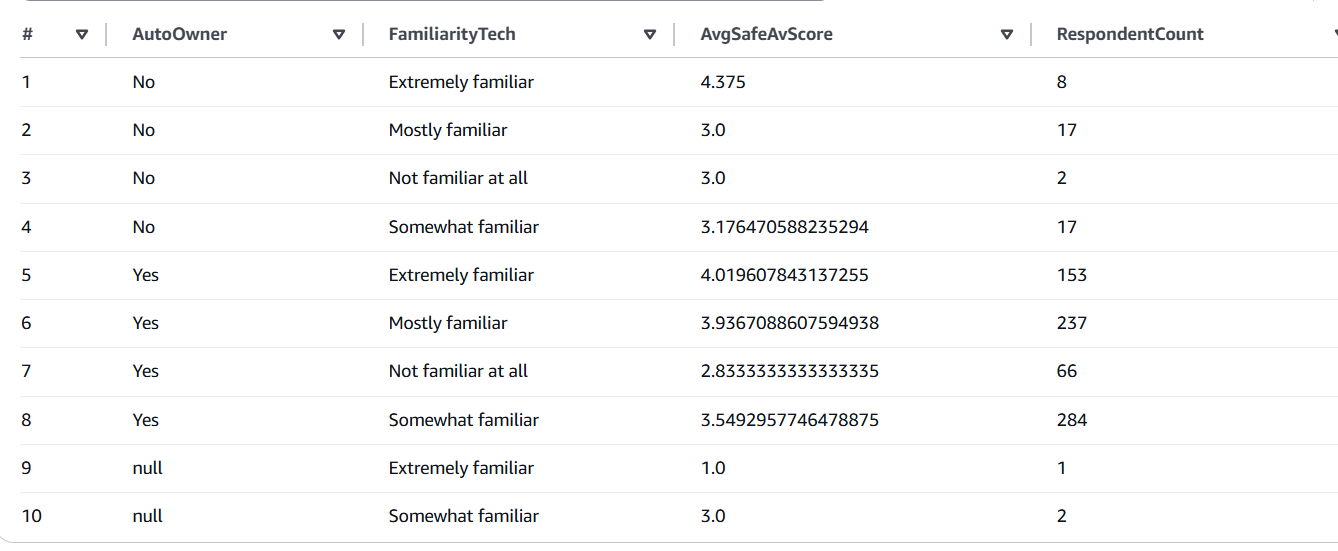}
    \caption{AV safety through tech familiarity and ownership}
    \label{fig:q5}
\end{figure}\textbf{}

\begin{figure}[h!]
    \centering
    \includegraphics[width=8cm]{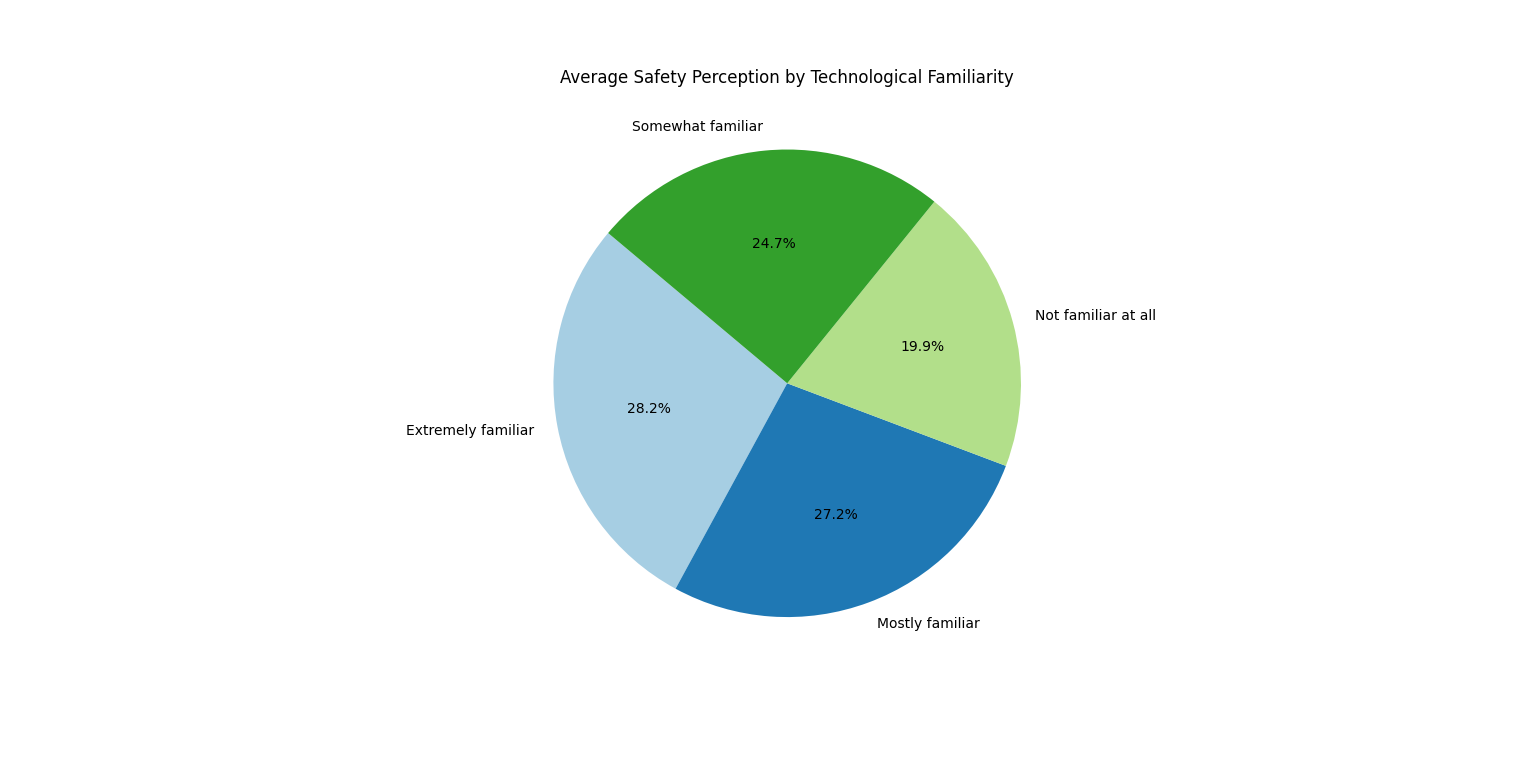}
    \caption{AV Safety and Technological Familiarity: The chart indicates that higher tech familiarity is associated with higher safety perceptions.}
    \label{fig:newfig}
\end{figure}

The 'AV Safety and Technological Familiarity' figure provides a clear illustration of the relationship between technological familiarity and autonomous vehicle safety perceptions. According to the statistics, people with a high level of technological familiarity are more likely than those with less exposure to technology to comprehend and trust the safety measures and procedures built into AV systems. This pattern implies that confidence in the potential and security of autonomous vehicles rises in tandem with technological literacy. In order to promote better acceptance and confidence in developing AV technologies, it emphasizes the necessity of activities targeted at improving public technological education.\\
In addition, the pie chart's distribution highlights a crucial component of consumer education and the uptake of new technology. Using these findings, stakeholders in the technology and automobile industries might create user engagement plans and educational programs that target particular demographics with less technological know-how. These initiatives have the potential to greatly enhance public awareness and open the door for broader adoption and use of autonomous vehicles by filling in the knowledge gap and demythologizing their capabilities. The chart is a tool for data visualization, but it also acts as a strategic roadmap for creating all-encompassing strategies to increase public confidence and safety guarantees in autonomous car technology..

\section{CONCLUSION}
This study has shed important light on how Pittsburgh residents view autonomous vehicles (AVs), emphasizing important elements that affect public confidence and adoption. According to the report, having specific infrastructure in place significantly improves the integration of AVs into urban environments. In addition to promoting cyclists' and pedestrians' physical safety, this infrastructure greatly increases their trust in the dependability of AVs' operations. Younger populations in particular show a greater inclination to trust AV technology, which is closely associated with their familiarity with AVs and level of technological literacy. The significance of coordinating AV development with strong educational programs that seek to demystify the technology and encourage widespread technical fluency is highlighted by this demographic trend. \\ 
Additional research highlights the necessity of changing policies to take into account how different demographics interact with AV technology. It is crucial to make sure that smart transportation solutions are safe and accessible for all users as cities develop and implement them. The findings of the study support a proactive strategy that puts safety and inclusion first in urban design and policy-making. Pittsburgh may set an example for the development of reliable and sustainable AV ecosystems by promoting an atmosphere where community needs and technology coexist. In addition to improving the daily urban experience, these initiatives are essential for establishing a standard for upcoming international urban transportation systems.

\section{Limitations}
It is important to take into account the many limitations of this study. Because some groups, such those who have little experience with shared infrastructure or autonomous vehicles, may be underrepresented, the dataset might not accurately reflect Pittsburgh's population's diversity. This might lead to results that don't accurately represent the general population. \\
Self-reported responses are the basis for many of the variables examined, including comfort levels, perceived safety, and willingness to share data. These are subjective by nature and may be impacted by personal prejudices, current events, or misinterpretations of the questions. The results could become inconsistent as a result.\\
Additionally, certain crucial elements that can affect perceptions are absent from the investigation, such as traffic patterns, psychological attitudes toward technology, or previous AV-related mishaps. The study's breadth is restricted by these missing variables, which also might be a factor in the regression model's poor prediction performance. \\
The nuanced variations in response levels may also be lost when ordinal variables, like technological familiarity, are reduced to numerical scores. Furthermore, the results only relate to Pittsburgh and could not be generalizable to other locations with varied infrastructure, traffic patterns, or AV adoption levels.\\
Finally, as autonomous vehicles develop and become more ingrained in daily life, public attitudes toward them are likely to shift. This study is a snapshot in time. By incorporating a larger population, more variables, and longitudinal data to offer more thorough insights, future study could overcome these constraints.

\section{Future Work}
Future research should think about broadening demographic and geographic coverage by gathering information from locations with varying infrastructure, traffic patterns, and degrees of autonomous vehicle (AV) adoption in order to improve on the findings of this study. Incorporating underrepresented groups can provide a more thorough knowledge of public attitude, particularly for individuals who have little exposure to shared infrastructure or AVs. In order to discover important experiences that affect acceptance and trust, longitudinal studies are especially crucial since they can monitor how public attitudes change over time as AV exposure increases.\\Furthermore, in order to better capture the intricate, non-linear interactions between demographic, behavioral, and contextual variables and produce more accurate predictions of AV acceptability, improved predictive modeling utilizing machine learning should be sought. Future studies should evaluate which communications techniques best foster confidence in AV technology, and public education campaigns should concentrate on enhancing technological literacy, particularly among communities with low levels of technological proficiency. Lastly, in order to assess how changing laws and regulations affect public perceptions of safety, accessibility, and justice, cooperation with legislators is essential. This will guarantee that AV deployment strategies continue to be inclusive, responsive, and based on empirical data.\\Improving data gathering instruments to better capture qualitative insights is another crucial avenue. Incorporating open-ended questions, interviews, and focus groups could reveal more profound psychological and emotional elements impacting public trust in AVs, even though organized surveys offer useful statistical data. Additionally, these qualitative approaches would offer complex viewpoints on edge instances or surprising encounters with AVs that would not be conveyed by conventional survey questions. Future research might also examine how sentiment analysis on social media might be used as an additional strategy to monitor changing public sentiment in real time.\\Lastly, researchers should think about adopting virtual or augmented reality environments to simulate different AV interaction scenarios. By testing human responses to AVs in controlled, reproducible settings, such immersive technologies can offer insightful behavioral data without posing real-world hazards. The public's reactions to various AV communication techniques, including signs, signals, or auditory cues, can also be assessed using these simulations. Future study on human-AV interaction will benefit from the use of such experimental designs, which will also help to support evidence-based advancements in safety procedures and AV interface design.

\end{document}